\documentclass[aps,pra,amsmath,amssymb,floatfix,twocolumn,amsmath,superscriptaddress,twocolumn,nofootinbib,tighten,letterpaper]{revtex4}
\usepackage{multirow}
\usepackage{subfigure}
\usepackage{color}
\usepackage{mathrsfs}
\usepackage{hyperref}
\usepackage[normalem]{ulem}
\usepackage{bm}

\usepackage{amssymb}   
\usepackage{amsmath}
\renewcommand\vec[1]{\ensuremath\boldsymbol{#1}} 

\usepackage{amsfonts, relsize, color}
\usepackage{graphics}
\usepackage{graphicx}
\usepackage{subfigure}
\usepackage{hyperref}
\usepackage{color}
\usepackage{comment}

\begin{document}

\title{Hierarchy of higher-order Floquet topological phases in three dimensions}

\author{Tanay Nag}
\affiliation{SISSA, via Bonomea 265, 34136 Trieste, Italy}
\affiliation{Institute f\" ur Theorie der Statistischen Physik, RWTH Aachen University, 52056 Aachen, Germany}

\author{Vladimir Juri\v ci\' c}
\affiliation{Nordita, KTH Royal Institute of Technology and Stockholm University, Roslagstullsbacken 23,  10691 Stockholm,  Sweden}
\affiliation{Departamento de F\' isica, Universidad T\' ecnica Federico Santa Mar\' ia, Casilla 110, Valpara\' iso, Chile}

\author{Bitan Roy}\thanks{Corresponding author: bitan.roy@lehigh.edu}
\affiliation{Department of Physics, Lehigh University, Bethlehem, Pennsylvania, 18015, USA}

\date{\today}

\begin{abstract}
Following a general protocol of periodically driving static first-order topological phases (supporting surface states) with suitable discrete symmetry breaking Wilson-Dirac masses, here we construct a hierarchy of higher-order Floquet topological phases in three dimensions.
In particular, we demonstrate realizations of both second-order and third-order Floquet topological states, respectively supporting dynamic hinge and corner modes at zero quasienergy, by periodically driving their static first-order parent states with one and two discrete symmetry breaking Wilson-Dirac mass(es). While the static surface states are characterized by codimension $d_c=1$, the resulting dynamic hinge (corner) modes, protected by \emph{antiunitary} spectral or particle-hole symmetries, live on the boundaries with $d_c=2$ $(3)$. We exemplify these outcomes for three-dimensional topological insulators and Dirac semimetals, with the latter ones following an arbitrary spin-$j$ representation.
\end{abstract}

\maketitle

\emph{Introduction}. The hallmark of topological phases of matter, the bulk-boundary correspondence, beyond the territory of static systems~\cite{hasan-kane:RMP, qi-zhang:RMP, chiu-teo-schnyder-ryu:RMP, armitage-mele-vishwanath:RMP, Shen-book, Bernevig-book}, is also operative on dynamic or driven quantum materials~\cite{galitski, gedik, moessner, berg, dseakdutta, eckardt, takashi, alu, azameit, Exp-1, Exp-2, Exp-3, Exp-4}. However, the bulk-boundary correspondence in dynamic systems is more subtle due to the nontrivial role of the time dimension. As such a static trivial phase can acquire nontrivial topology under suitable periodic drive, for example. Due to the time translational symmetry, the resulting Floquet topological phase then features steady state or nondissipative topological modes, localized at the boundaries.

\begin{figure}[t!]
\includegraphics[width=0.80\linewidth]{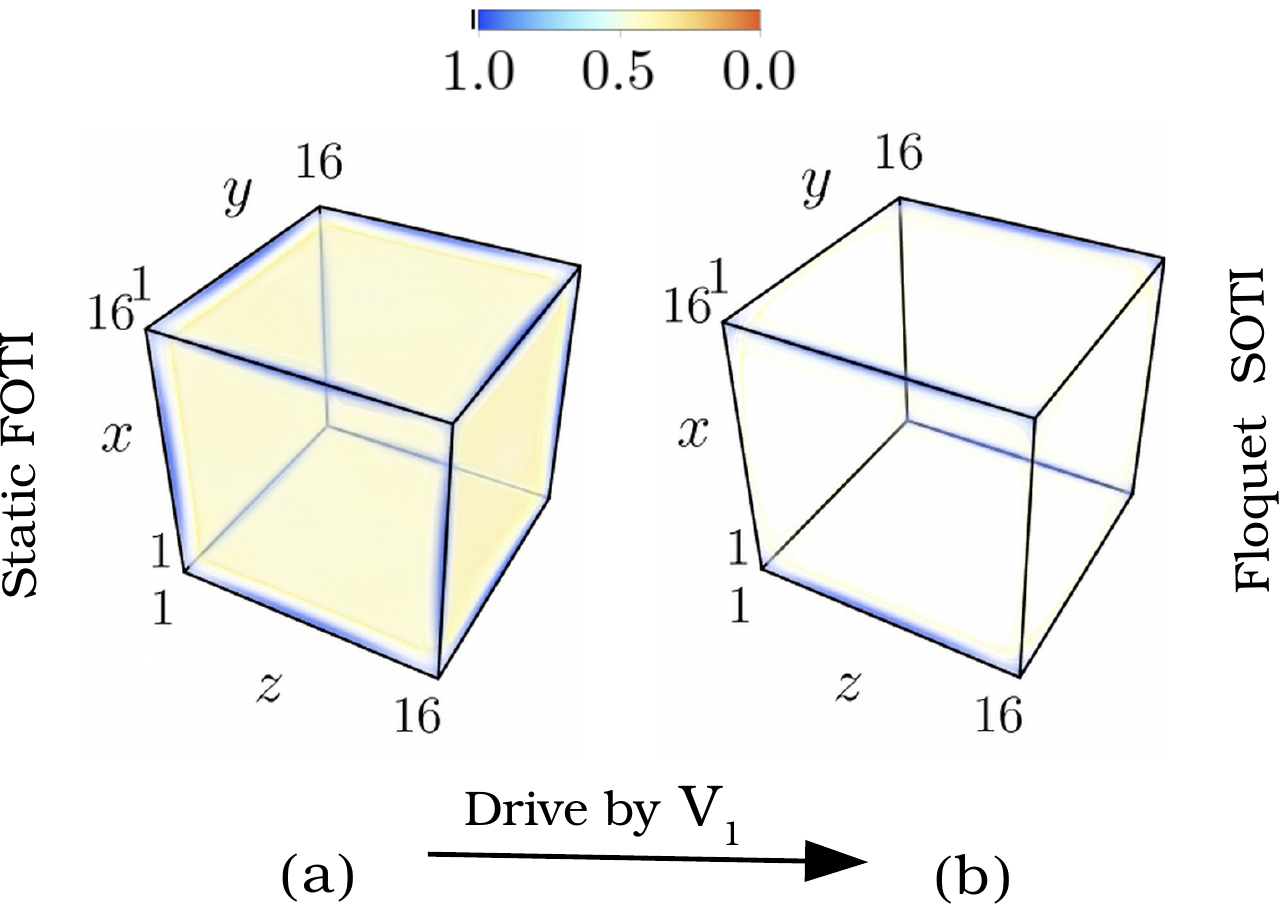}
\caption{(a) LDOS for the zero energy surface states of a static FOTI for $t=2t_0=1$ and $m=1$ [Eq.~\eqref{Eq:2TI_Hamiltonian} involving four-dimensional $\Gamma$ matrices]. (b) LDOS associated with one-dimensional Floquet hinge modes with zero quasienergy after periodically kicking a static FOTI with the $C_4$ symmetry breaking Wilson-Dirac mass term [Eq.~\eqref{Eq:periodickick3}] for $\Delta_1=0.17$ and drive frequency $\omega (=2\pi/T)=10 \gg t,t_0$ (ensuring the high-frequency regime). Notice that the hinge quasimodes are localized along the $C_4$ rotational symmetry breaking $z$ axis. In both cases, the LDOS is normalized by its maximal value.}~\label{fig:2nd_TI}
\end{figure}

\begin{figure}[t!]
\includegraphics[width=0.95\linewidth]{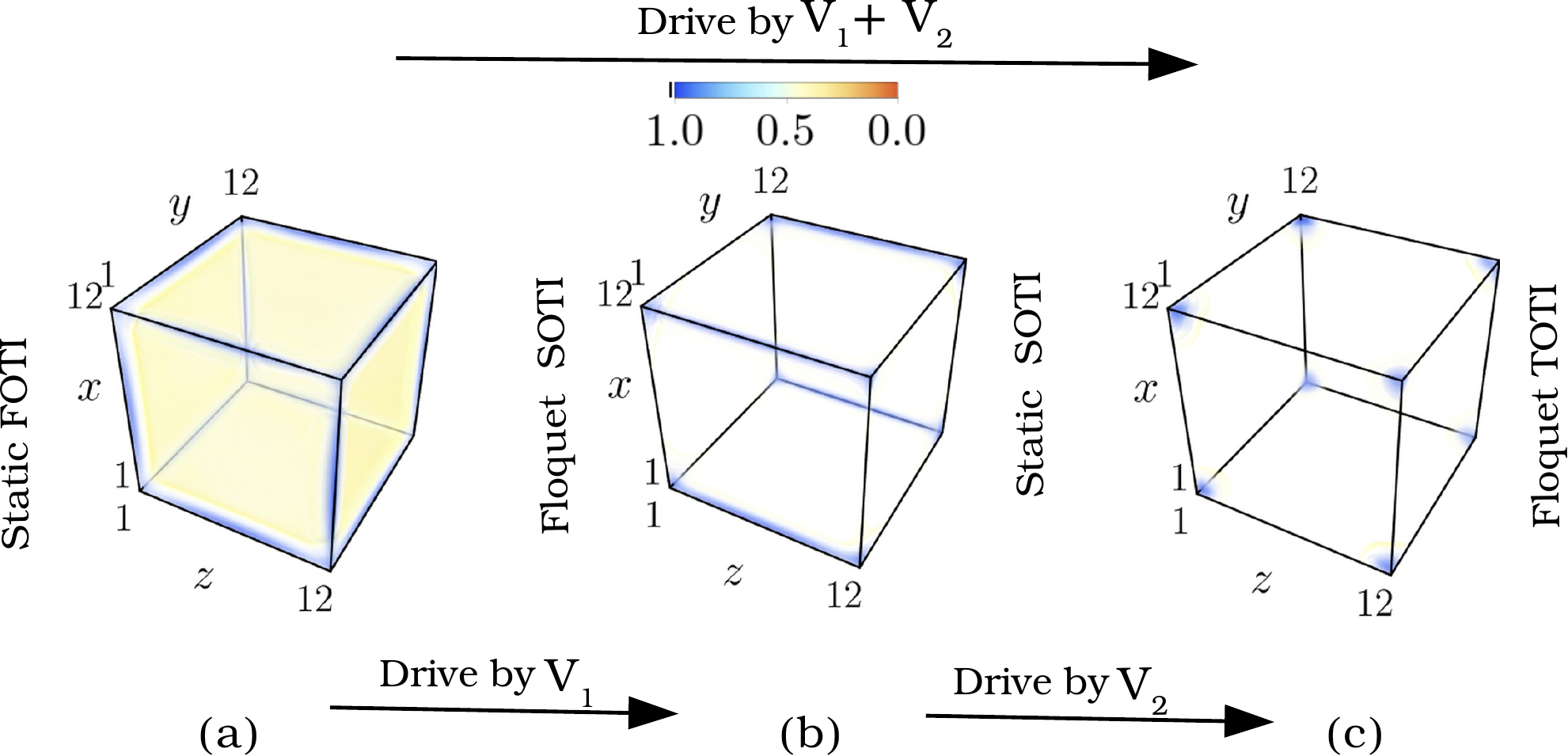}
\caption{LDOS for (a) surface states of a static FOTI (involving eight-dimensional $\Gamma$ matrices), (b) dynamic hinge modes of Floquet SOTI (by periodically driving a static FOTI by $V_1$), and (c) dynamic corner modes of Floquet TOTI (by periodically driving a static FOTI by $V_1$ and $V_2$), for $t=t_0=1$, $m=2$, $\Delta_1=\Delta_2=0.3$, and $\omega=10\gg t,t_0$ (assuring the high-frequency regime), see Eq.~(\ref{Eq:periodickick4}). We also find dynamic corner modes as in (c), when a \emph{static} SOTI is periodically driven by $V_2$. Throughout LDOS is normalized by its maximum value.
}
\label{fig:2nd_TI_8band}
\end{figure}

\begin{figure}[t!]
\includegraphics[width=0.95\linewidth]{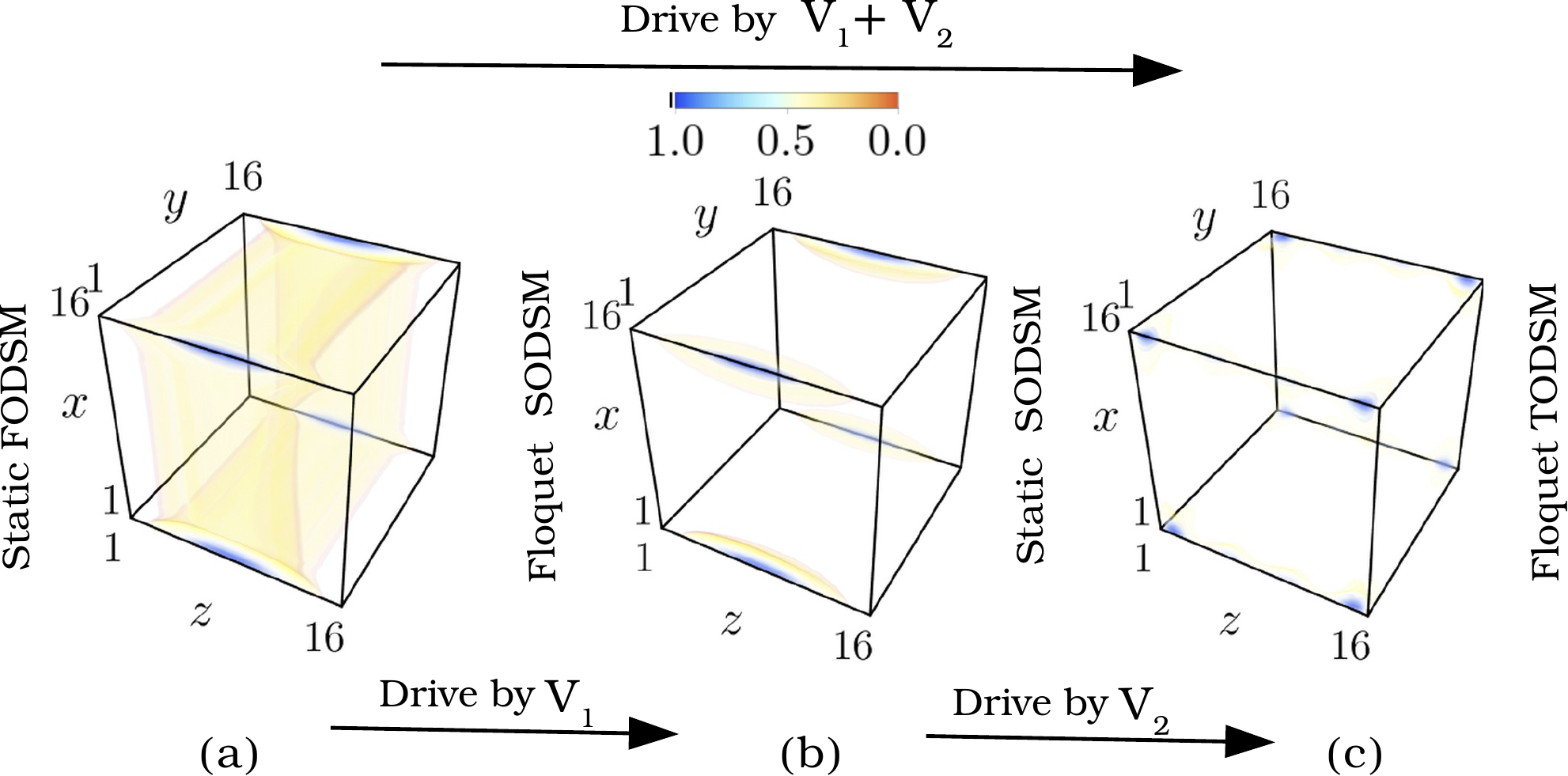}
\caption{LDOS associated with (a) zero energy surface Fermi arc states of a FODSM, (b) zero quasienergy hinge modes in a Floquet SODSM, and (c) zero quasienergy corner modes in a Floquet TODSM, respectively obtained by periodically driving the FODSM with only $V_1$, and $V_1$ and $V_2$, in a spin-1/2 Dirac system. Here we set $t=t_z=m=1$, $\Delta_1=0.3$, $\Delta_2=0.3$ and $\omega=10$ (ensuring the high-frequency regime) in Eqs.~\eqref{Eq:DSM_Hamiltonian} and~\eqref{Eq:periodickick}. One can also obtain zero quasienergy corner modes as in (c), by periodically driving a \emph{static} SODSM, supporting four zero energy hinge modes, with $V_2$. Throughout LDOS is normalized by its maximum value.
}~\label{fig:DSM}
 \end{figure}

\begin{figure}[t!]
\includegraphics[width=0.95\linewidth]{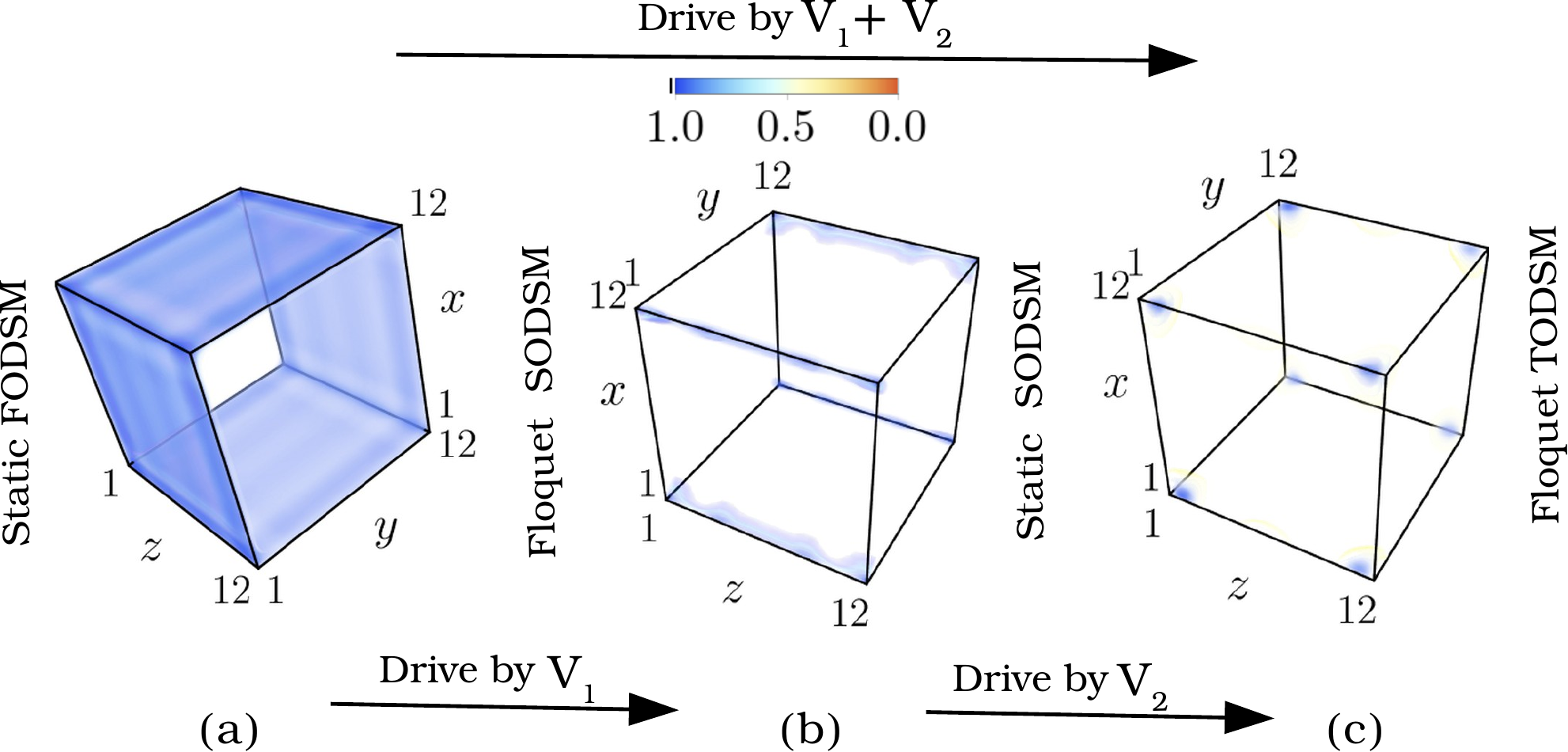}
\caption{Same as Fig.~\ref{fig:DSM}, but for spin-1 Dirac systems.}~\label{fig:DSM-spin-1}
 \end{figure}

Typically, a $d$-dimensional topological phase supports boundary modes on $(d-1)$-dimensional interface with vacuum, also characterized by the \emph{codimension} $d_c=d-(d-1)=1$. The known examples are the edge (surface) states of two- (three-) dimensional topological insulators and semimetals~\cite{hasan-kane:RMP, qi-zhang:RMP, chiu-teo-schnyder-ryu:RMP, armitage-mele-vishwanath:RMP, Shen-book, Bernevig-book}. The concept of the bulk-boundary correspondence has been recently generalized to topologically protected modes living on boundaries of codimension $d_c=n>1$, with $n$ as an integer~\cite{BBH-Science, BBH-PRB, Langbehn-PRL2017, Schindler-SciAdv2018, Hsu-PRL2018, Liu-Hughes-PRB2018, Trifunovic-PRX2019, calugaru-juricic-roy, Varjas-PRL2019, agarwala-PRR2020, roy-antiunitary, fang-fu:sciadv, zeng-PRB-2020, andras-2019, bernevig:HOTDSM, DasSarma-arxiv2019, andras2020, murakami:arXiv2020, wangwang:arXiv2020, ghorashi-arxiv2020, jiang-arxiv2020, watanabe:arxiv2020}. Frequently encountered examples of such lower-dimensional boundary modes in so-called higher-order topological states are the corner and hinge modes, respectively characterized by $d_c=d$ and $d-1$~\cite{codimension}. Although the generalized bulk-boundary correspondence has been extended to driven or Floquet setups~\cite{Gong2019, Seradjeh2018, diptiman2019, Gilrefael2019, Plekhanov2019, Nag2019, zhao-liu2020, huang-liu2020, ghosh-paul-saha2020, bomantara2020, pan-zhou2020, zhang-yang-fragileFloquet, wu-wang-an2020}, it somewhat exclusively focuses on two-dimensional systems (see, however, Ref.~\cite{zhao-liu2020}), supporting only Floquet corner modes. Here we show how one can systematically realize three-dimensional (3D) Floquet second-order and third-order phases, respectively supporting dynamic hinge and corner modes at zero quasienergy [Figs.~\ref{fig:2nd_TI}-~\ref{fig:DSM-spin-1}], by periodically driving static first-order topological phases (both insulators and semimetals) with discrete rotational symmetry breaking Wilson-Dirac masses.

We now present a summary of our main results. We show that when periodically driven by a four-fold ($C_4$) symmetry breaking Wilson-Dirac mass, a 3D static first-order topological insulator (FOTI), supporting surface states, can be converted into a Floquet second-order topological insulator (SOTI), accommodating four one-dimensional hinge modes along the symmetry breaking $z$ axis, for example, see Fig.~\ref{fig:2nd_TI}. As the parent 3D static FOTI involves four mutually anticommuting Hermitian matrices, one can realize only a Floquet SOTI when the $\Gamma$ matrices are four-dimensional [Eqs.~(\ref{Eq:2TI_Hamiltonian}) and (\ref{Eq:periodickick3})]. On the other hand, an eight-dimensional representation of the $\Gamma$ matrices, thus permitting additional anticommuting mass matrices [Eq.~(\ref{Eq:gammatriceseightdimensional})], when accompanied by a suitable discrete symmetry breaking momentum-dependent form factor, facilitates realization of a Floquet third-order topological insulator (TOTI), supporting eight zero quasienergy corner modes, see Fig.~\ref{fig:2nd_TI_8band}. By contrast, a 3D first-order Dirac semimetal (FODSM), supporting Kramers degenerate Fermi arc surface states and following arbitrary spin-$j$ representation, can always be augmented by two discrete symmetry breaking masses. They allow realizations of both Floquet second-order Dirac semimetal (SODSM) and third-order Dirac semimetal (TODSM), respectively supporting one-dimensional hinge and pointlike corner states at zero quasienergy. We explicitly demonstrate these outcomes for spin-1/2 [Fig.~\ref{fig:DSM}] and spin-1 [Fig.~\ref{fig:DSM-spin-1}] Dirac systems. Importantly, the dynamic hinge and corner modes are always pinned at zero quasienergy by an \emph{antiunitary} particle-hole symmetry~\cite{roy-antiunitary, Nag2019}.

\emph{Floquet HOTI.} First we demonstrate generation of a dynamical higher-order topological insulator (HOTI) within the Floquet framework starting from a \emph{static} FOTI, described by the Hamiltonian~\cite{sczhang-natphys, juricic-natphys}
\allowdisplaybreaks[4]
\begin{eqnarray}~\label{Eq:2TI_Hamiltonian}
H^{\rm stat}_{\rm FOTI} &=& t \sum_{j=1}^3\Gamma_j \; S_j + \Gamma_4 \; [(m-6 t_0)+ 2 t_0 \sum^3_{j=1}C_j] \nonumber \\
&\equiv& \sum^{4}_{j=1} N_j (\vec{k}) \; \Gamma_j,
\end{eqnarray}
with $S_j\equiv\sin(k_ja)$, $C_j\equiv\cos(k_ja)$, and $\vec{k}$ as momentum. We set the lattice spacing $a=1$ hereafter. The mutually anticommuting four-component $\Gamma$ matrices are $\Gamma_i=\sigma_1  \tau_i$, for $i=1,2,3$, and $\Gamma_4= \sigma_3  \tau_0$. The Pauli matrices $\tau_\mu$ ($\sigma_\mu$) operate on the orbital (spin) degrees of freedom. For $0<\frac{m}{t_0}<4$ the band inversion takes place at the $\Gamma=(0,0,0)$ point of the Brillouin zone. The resulting topological surface states get pinned at zero energy due to both unitary and antiunitary spectral or particle-hole symmetries, generated by the operators $C=\Gamma_5$, with $\Gamma_5= \sigma_2  \tau_0$, and $P= \sigma_2 \tau_2 {\mathcal K}$, respectively, since $\{ H^{\rm stat}_{\rm FOTI}, C \}=\{ H^{\rm stat}_{\rm FOTI}, P \}=0$, where ${\mathcal K}$ is the complex conjugation~\cite{antiunitarymomentum, terminology}. To generate a Floquet SOTI we periodically kick the FOTI by the HOT mass term
\begin{equation}~\label{Eq:periodickick3}
V(t) =  V_1 \; \Gamma_{5} \sum^{\infty}_{r=1} \; \delta \left( t- r \; T \right),
\end{equation}
where $V_{1}= \sqrt{3} \Delta_1 \left( \cos k_1-\cos k_2 \right)$ is the discrete $C_4$ symmetry breaking Wilson-Dirac mass and $T$ is the kicking period. In a static system $V_1 \Gamma_5$ term gaps out otherwise gapless surface modes of the FOTI on both $xz$ and $yz$ surfaces leaving only their four intersections, where $V_1$ changes its sign, gapless. It is due to the fact that $V_1$ vanishes along $k_x=\pm k_y$ for any $k_z$ in the momentum space and when we convert $V_1$ into real space hopping, it changes sign across the diagonals $x=\pm y$ for any $z$, yielding the zero energy hinge modes in the $z$ direction. As a consequence, the static SOTI, described by the Hamiltonian $H^{\rm stat}_{\rm SOTI}=H^{\rm stat}_{\rm FOTI} + V_1 \Gamma_5$, features one-dimensional gapless propagating hinge modes along the $C_4$ symmetry breaking $z$ axis. By contrast, the $xy$ surfaces continue to host gapless Dirac states, as $V_1$ vanishes at the center of the corresponding surface Brillouin zone $(k_x,k_y)=(0,0)$, where the apex of the surface Dirac cone is pinned at. Next we show when a static FOTI [Eq.~\eqref{Eq:2TI_Hamiltonian}] is periodically kicked by such a discrete $C_4$ symmetry breaking Wilson-Dirac mass [Eq.~\eqref{Eq:periodickick3}], it dynamically generates a Floquet SOTI, supporting one-dimensional Floquet hinge modes.

To this end, we compute the Floquet operator associated with the Hamiltonian $H^{\rm stat}_{\rm FOTI}+V(t)$, given by
\begin{align}~\label{Eq:Floquetoperator}
U(\vec{k},T) &= {\rm TO}\left( \exp \left[ -i\int_0^T \left[H^{\rm stat}_{\rm FOTI} + V(t) \right] dt \right] \right)  \nonumber \\
     &= \exp(-i H^{\rm stat}_{\rm FOTI} \; T) \; \exp(-i   V_{1} \; \Gamma_5 ),
\end{align}
where `$\rm TO$' stands for the time-ordered product. After a single kick, the Floquet operator
assumes the form
\begin{equation}~\label{Eq:Floquet3}
U(\vec{k},T) = C_T \left[ n_0 -i n_5 \Gamma_5 \right] - i S_T \sum^4_{j=1} \left[ n_j \Gamma_j + m_{j} \Gamma_{j5} \right],
\end{equation}
where $\Gamma_{jk}=[\Gamma_j, \Gamma_k]/(2i)$, $C_T= \cos (|{\bf N}(\vec{k})| T)$, $S_T= \sin (|{\bf N}(\vec{k})| T)$, $n_0 = \cos (V_1)$, $n_5 = \sin(V_1)$, and $(n_j,m_j)= N_j (\vec{k}) \; (n_0,n_5)/|{\bf N} (\vec{k})|$ for $j=1, \cdots, 4$. The effective Floquet Hamiltonian $H_{\rm Flq}=i \ln (U(\vec{k},T))/T$ reads
\begin{equation}~\label{Eq:non-pert-FloquetHam}
H_{\rm Flq}=\frac{\epsilon_{\vec{k}}}{\sin(\epsilon_{\vec{k}}T)}\bigg[S_T  \sum_{j=1}^{4} \left( n_j \; \Gamma_j + m_j \; \Gamma_{j5} \right) + C_T \; n_5 \; \Gamma_5 \bigg]
\end{equation}
with $\epsilon_{\bm k}=\arccos (C_T n_0)/T$, which in the high-frequency limit ($T \to 0$, $V_1 \to 0$, but finite $V_1/T$) takes the form
\begin{equation}~\label{Eq:FloquetHamiltonian-TI}
H^{\rm HF}_{\rm Flq} = \sum^4_{j=1} N_j ({\bf k}) \Gamma_j + V_{1} \; \sum^4_{j=1} N_{j} ({\bf k}) \Gamma_{j5} + \frac{V_1}{T} \; \Gamma_{5}.
\end{equation}
Therefore, the effective Floquet Hamiltonian ($H_{\rm Flq}$ or $H^{\rm HF}_{\rm Flq}$) only preserves the antiunitary particle-hole symmetry, since $\{ H_{\rm Flq}, P \}=0=\{ H^{\rm HF}_{\rm Flq}, P \}$, which in turn pins the dynamical hinge modes at zero quasienergy.

To show the hallmark hinge modes of the Floquet SOTI, next we numerically solve for the zero quasienergy states of the Floquet operator in Eq.~(\ref{Eq:Floquetoperator}), satisfying
\begin{equation}~\label{Eq:quasienergy_def}
U(\vec{k},T) \; |\phi_n \rangle =\exp(i \mu_n T) \; |\phi_n \rangle
\end{equation}
on a cubic lattice with open boundaries in all three directions. Here, $|\phi_n \rangle$ is the Floquet state with quasienergy $\mu_n$. In Fig.~\ref{fig:2nd_TI}(b), we display the local density of states (LDOS) associated with the (almost) zero [${\mathcal O} (10^{-6})$] quasienergy Floquet states, starting from a parent FOTI, supporting surface states [Fig.~\ref{fig:2nd_TI}(a)]. Therefore, when a static FOTI is periodically kicked by a mass term, breaking the $C_4$ rotational symmetry about the $z$ axis, the one-dimensional zero quasienergy Floquet hinge modes, guaranteed by the antiunitary particle-hole symmetry $P^{-1}U(\vec{k},T)P=U(\vec{k},T)$, appear along the symmetry breaking $z$ axis. Analogously, the breaking of the $C_4$ rotational symmetry about the $x$ or $y$ axis results in the hinge modes along the same axis.

Notice that four-dimensional Hermitian matrices accommodate maximal \emph{five} mutually anticommmuting matrices. We have exhausted all of them to generate a static or Floquet SOTI. Therefore, we cannot proceed further to explore the hierarchy of HOTIs and construct a TOTI, supporting \emph{corner} modes, by partially gapping out the $z$ directional hinge and $xy$ surface modes. This constraint is removed when the $\Gamma$ matrices are \emph{eight-dimensional}, which, on the other hand, sustain \emph{seven} mutually anticommuting matrices. For concreteness, we commit to the following representation of seven mutually anticommuting eight-dimensional Hermitian $\Gamma$ matrices
\begin{align}~\label{Eq:gammatriceseightdimensional}
\Gamma_1&=\Sigma_1 \sigma_1  \tau_1, \Gamma_2=\Sigma_1 \sigma_1  \tau_2, \Gamma_3= \Sigma_1\sigma_1  \tau_3, \Gamma_4= \Sigma_1 \sigma_3  \tau_0, \nonumber \\
\Gamma_5&=\Sigma_1 \sigma_2  \tau_0, \;\; \Gamma_6=\Sigma_3 \sigma_0 \tau_0, \;\; \Gamma_7=\Sigma_2 \sigma_0 \tau_0.
\end{align}
The newly introduced Pauli matrices $\Sigma_\mu$ operate on the sublattice degrees of freedom, for example. Next we periodically drive such a FOTI by \emph{two} discrete symmetry breaking Wilson-Dirac masses
\begin{align}~\label{Eq:periodickick4}
V(t) =  \bigg( V_1  \; \Gamma_{5} + V_2 \; \Gamma_6 \bigg) \: \sum^{\infty}_{r=1} \; \delta \left( t- r \; T \right),
\end{align}
where $V_2= \Delta_2 (2 \cos k_3 - \cos k_1 -\cos k_2)$.

Before delving into the topology of such driven system, let us ignore the time dependence of two Wilson-Dirac masses and consider the following static Hamiltonian
\begin{equation}~\label{eq:3dcornerHamil}
H^{\rm stat}_{\rm TOTI}= H^{\rm stat}_{\rm FOTI} + V_1  \; \Gamma_{5} + V_2 \; \Gamma_6 \equiv H^{\rm stat}_{\rm SOTI} + V_2 \; \Gamma_6.
\end{equation}
Recall that the spectrum of $H^{\rm stat}_{\rm SOTI}=H^{\rm stat}_{\rm FOTI} + V_1 \Gamma_{5}$ (for $\Delta_2=0$) supports four gapless hinge modes along the $z$ direction and gapless surface states on the $xy$ planes, where $V_1$ vanishes. The second Wilson-Dirac mass $V_2$ vanishes only along eight body-diagonal $(\pm 1, \pm 1, \pm 1)$ directions, when simultaneously present with $V_1$~\cite{comment1}. Hence, $V_2$ further gaps out the gapless modes of $H^{\rm stat}_{\rm SOTI}$, leaving only eight corners of the cubic system gapless. As a result $H^{\rm stat}_{\rm TOTI}$ supports eight corner modes and we then realize a static TOTI. This conclusion can be further corroborated by the fact that the above Hamiltonian is equivalent to the 3D Benalcazar-Bernevig-Hughes (BBH) model, also featuring eight corner modes~\cite{BBH-Science}. Namely, for a specific choice of parameters $\Delta_1=\Delta_2=\sqrt{2} t_0$, the Hamiltonian in Eq.~(\ref{eq:3dcornerHamil}) maps onto the BBH model, given by $H^{\rm BBH}_{\rm TOTI}=H_1+H_2+H_3$, where for $j=1,2,3$
\begin{eqnarray}
H_j = \left[ \lambda+ t_1 \cos(k_j) \right] \gamma_{2 j} + t_1 \sin (k_j) \gamma_{2 j-1},
\end{eqnarray}
and $\gamma_j$ are mutually anticommuting eight-dimensional Hermitian matrices, satisfying $\{\gamma_j, \gamma_k \}=2 \delta_{jk}$, describing an octupolar insulator for $|\lambda/t_1|<1$. The mapping between $H^{\rm stat}_{\rm TOTI}$ and $H^{\rm BBH}_{\rm TOTI}$ is then set by $t_1=2\sqrt{3} t_0$, $\lambda=(m-6 t_0)/\sqrt{3}$, and
\allowdisplaybreaks[4]
\begin{align}
\gamma_1&=\Gamma_1, \; \gamma_2 = \frac{\Gamma_4}{\sqrt{3}} + \frac{\Gamma_5}{\sqrt{2}}-\frac{\Gamma_6}{\sqrt{6}}, \; \gamma_3=\Gamma_3,
 \nonumber \\
 \gamma_4 &=\frac{\Gamma_4}{\sqrt{3}} - \frac{\Gamma_5}{\sqrt{2}}-\frac{\Gamma_6}{\sqrt{6}}, \; \gamma_5=\Gamma_5, \;
 \gamma_6 = \frac{\Gamma_4}{\sqrt{3}}-\sqrt{\frac{2}{3}} \; \Gamma_6.
\end{align}

The corner modes are pinned at zero energy due to both unitary and antiunitary particle-hole symmetry of $H^{\rm stat}_{\rm TOTI}$, respectively generated by $C=\Gamma_7$ [Eq.~\eqref{Eq:gammatriceseightdimensional}] and $P=\Sigma_1 \sigma_2 \tau_2 {\mathcal K}$, as $\{ H^{\rm stat}_{\rm TOTI}, C \}=0= \{ H^{\rm stat}_{\rm TOTI}, P\}$. Next we show that this mechanism is also operative in the dynamical realm: when a static FOTI is periodically driven by two Wilson-Dirac masses [Eq.~\eqref{Eq:periodickick4}], one dynamically generates eight Floquet corner modes at zero quasienergy, and concomitantly a Floquet TOTI.

The effective Floquet Hamiltonian is obtained from the corresponding Floquet operator $U(\vec{k},T)=\exp(-i H^{\rm stat}_{\rm FOTI} \; T) \exp(-i V_1 \Gamma_5 -i V_2 \Gamma_{6})$, which in the high-frequency limit reads as
\begin{align}~\label{Eq:FloquetHamiltonian2}
H^{\rm HF}_{\rm Flq} = \sum^4_{j=1} N_j (\vec{k}) \left[  \Gamma_j +  V_1 \Gamma_{j5} +  V_2 \Gamma_{j6} \right]
+ \sum_{j=1}^2\frac{V_{j}}{T} \; \Gamma_{j+4}.
\end{align}
Its spectral symmetry is guaranteed only by the antiunitary operator $P=\Sigma_1 \sigma_2 \tau_2 {\mathcal K}$. By diagonalizing the Floquet operator $U(\vec{k}, T)$ on a cubic lattice with open boundaries, we find eight sharp corner localized modes at zero quasienergy, since $P^{-1}U(\vec{k},T)P=U(\vec{k},T)$, the hallmark of a Floquet TOTI, see Fig.~\ref{fig:2nd_TI_8band}.

Alternatively, one can start with a static SOTI, described by the Hamiltonian $H^{\rm stat}_{\rm SOTI}$, and periodically drive the system with the second Wilson-Dirac mass ($V_2$). The corresponding Floquet operator $U(\vec{k},T)= \exp(-i H^{\rm stat}_{\rm SOTI} \; T) \exp(-i V_2 \Gamma_{6})$ yields an effective Floquet Hamiltonian, which in the high-frequency limit reads
\begin{eqnarray}~\label{Eq:FloquetHamiltonian_TI3}
H^{\rm HF}_{\rm Flq} = \sum^5_{j=1} N_j (\vec{k}) \Gamma_j + V_{2} \; \sum^5_{j=1} N_{j} (\vec{k}) \Gamma_{j6} + \frac{V_2}{T} \; \Gamma_{6},
\end{eqnarray}
with $N_5(\vec{k}) \equiv V_1$. Once again its spectral symmetry is guaranteed by the antiunitary operator $P=\Sigma_1 \sigma_2 \tau_2 {\mathcal K}$, and $P^{-1}U(\vec{k},T)P=U(\vec{k},T)$. By diagonalizing the corresponding Floquet operator $U(\vec{k},T)$, we find eight sharp corner localized modes with zero quasienergy and a Floquet TOTI from a static SOTI [Fig.~\ref{fig:2nd_TI_8band}]. Following Refs.~\cite{hughes:multipole, cho:multipole}, we compute the octupole moment ($Q_{xyz}$) from $H^{\rm HF}_{\rm Flq}$ appearing in Eqs.~(\ref{Eq:FloquetHamiltonian2}) and (\ref{Eq:FloquetHamiltonian_TI3}), and obtain $Q_{xyz}=0.5$ for the TOTI, after subtracting the contribution in the atomic limit~\cite{agarwala-PRR2020}.

\emph{Floquet HODSM}. We now study 3D higher-order Dirac semimetals (HODSMs). A static FODSM, following the spin-$j$ representation, is described by the Hamiltonian
\begin{equation}~\label{Eq:DSM_Hamiltonian}
H^{\rm stat}_{\rm FODSM}= \sum^3_{j=1} N_j ({\bm k}) \; \Gamma_j,
\end{equation}
where $N_j(\vec{k})= t \sin (k_j )$ for $j=1,2$, and $N_3 (\vec{k})=t_z \cos k_3 + m (\cos k_1+\cos k_2 -2)$. The $2(2j+1)$-dimensional $\Gamma$ matrices are $\Gamma_1=S_1 \tau_3$, $\Gamma_2=S_2 \tau_0$, $\Gamma_3=S_3 \tau_0$, where $\vec{S}$ are the spin-$j$ matrices. Only for spin-1/2 systems, with $\vec{S}\equiv {\boldsymbol \sigma}$, three $\Gamma$ matrices mutually anticommute. Nevertheless, for any half integer $j$, the valence and conduction bands touch at two Dirac points located at $\vec{k}_\star=(0,0,\pm \pi/2)$ when $t=t_z=m=1$, yielding $(2j+1)$ copies of linearly dispersing bands in their vicinity. On the other hand, for an integer $j$, besides $2j$ linearly dispersing bands the system supports a trivial flat band at zero energy. Irrespective of the value of $j$, a FODSM accommodates $2 j$ copies of Kramers degenerate Fermi arc surface states~\cite{calugaru-juricic-roy, nandy-manna-calugaru-roy}. In the real space they occupy $xz$ and $yz$ surfaces, as shown in Figs.~\ref{fig:DSM} and~\ref{fig:DSM-spin-1}, respectively for $j=1/2$ and $j=1$. The Fermi arc states are pinned at zero energy by the antiunitary particle-hole symmetry, generated by $P=M_{ad} \; \tau_3 {\mathcal K}$, where $M_{\rm ad}$ is the $(2j+1)$-dimensional \emph{antidiagonal} matrix. To dynamically generate the HODSM, we periodically drive the FODSM by the discrete symmetry breaking Wilson-Dirac masses
\begin{align}~\label{Eq:periodickick}
V(t) = \sum_{j=1}^2  V_j \; \Gamma_{3+j} \sum^{\infty}_{r=1} \; \delta \left( t- r \; T \right),
\end{align}
where $V_{1}= \Delta_1 \left( \cos k_1-\cos k_2 \right)$, $V_2= \Delta_2 \sin ( 2 k_3)$, $\Gamma_4=M_{\rm ad} \; \tau_1$ and $\Gamma_5=M_{\rm ad} \; \tau_2$.

Neglecting the time dependence of $V(t)$, we first focus on the static Hamiltonian $H^{\rm stat}_{\rm HODSM}=H^{\rm stat}_{\rm FODSM}+
V_1 \Gamma_4 + V_2 \Gamma_5$, which describes a static SODSM (TODSM) for $|\Delta_1| \neq 0$ and $\Delta_2=0$ ($|\Delta_1|, |\Delta_2| \neq 0$). Notice that $V_1$ ($V_2$) represents quadrupolar (dipolar) order, but both of them vanish at the Dirac points $\vec{k}_\star$, leaving them unaffected. In particular, as in the case of the SOTI, the Wilson-Dirac mass $V_1 \Gamma_4$ yields a SODSM with four hinge states in the $z$-direction~\cite{Liu-Hughes-PRB2018, calugaru-juricic-roy, bernevig:HOTDSM, andras2020}. The term $V_2 \Gamma_5$ then acts as a mass domain wall for these hinge states and gaps them out everywhere except at the eight corners, yielding a TODSM. Both the hinge and corner modes are pinned at zero energy as $\{ H^{\rm stat}_{\rm HODSM}, P \}=0$. We now periodically drive a FODSM with $V(t)$ [Eq.~\eqref{Eq:periodickick}] and generate the Floquet SODSM and TODSM, respectively with hinge and corner modes at zero quasienergy.

The corresponding Floquet operator $U(\vec{k},T)=\exp(-i H^{\rm stat}_{\rm FODSM} T) \exp(-i V_1 \Gamma_4 -i V_2 \Gamma_5)$, satisfies the antiunitary particle-hole symmetry $P^{-1}U(\vec{k},T)P=U(\vec{k},T)$ for arbitrary $\Delta_1$, $\Delta_2$, and $j$. Indeed by diagonalizing this Floquet operator on a cubic lattice with open boundaries we find (1) a SODSM with four zero quasienergy hinge modes for $\Delta_2=0$ and (2) a TODSM with zero quasienergy corner modes when both $\Delta_1$ and $\Delta_2$ are finite. Even though our results hold for arbitrary value of $j$, here the results are displayed for only $j=1/2$ and $j=1$ in Figs.~\ref{fig:DSM} and~\ref{fig:DSM-spin-1}, respectively. Only for $j=1/2$ the effective Floquet Hamiltonian can be written compactly, which in the high-frequency limit reads
\begin{align}~\label{Eq:FloquetHamiltonian}
H^{\rm HF}_{\rm Flq} = \sum^3_{j=1} N_j (\vec{k}) \left[ \Gamma_j + V_{1} \Gamma_{j4} + V_{2} \Gamma_{j5} \right]
+ \sum_{j=1}^2\frac{V_{j}}{T} \; \Gamma_{j+3},
\end{align}
as $T, \Delta_{1,2} \to 0$, but $\Delta_{1,2}/T$ is finite, and preserves the antiunitary particle-hole symmetry as $\{H^{\rm HF}_{\rm Flq}, P \}=0$.

Alternatively, one can also start with a static SODSM, described by the Hamiltonian $H^{\rm stat}_{\rm SODSM}=H^{\rm stat}_{\rm FODSM} + V_1 \Gamma_4$, supporting four zero energy hinge modes, and periodically drive the system with the second Wilson-Dirac mass $V_2 \Gamma_5$. The corresponding Floquet operator $U(\vec{k},T)=\exp(-i H^{\rm stat}_{\rm SODSM} \; T) \exp(-i V_2 \Gamma_5)$ satisfies $P^{-1}U(\vec{k},T)P=U(\vec{k},T)$. By diagonalizing $U(\vec{k},T)$ on an open cubic lattice we find zero quasienergy corner modes, the hallmark of a TODSM, see Figs.~\ref{fig:DSM} and~\ref{fig:DSM-spin-1}. At least for the spin-1/2 SODSM, we find that the quadrupole moment~\cite{hughes:multipole, cho:multipole} $Q_{xy} (k_z)=0.5$ after subtracting the contribution in the atomic limit~\cite{agarwala-PRR2020} between two Dirac points, i.e. for $|k_z| \leq |\vec{k}_\star|$, when computed from the high-frequency effective Hamiltonian.

\emph{Summary and discussion}. Here we show that starting from a first-order topological phase (insulator or semimetal), one can systematically explore the cascade of dynamic HOT phases in three dimensions, when periodically driven by suitable discrete symmetry breaking Wilson-Dirac masses~\cite{intrinsicextrinsic:comment}. Specifically, we show that in the presence of a single (two) dynamic mass(es) one can realize second-order (third-order) Floquet topological phases, supporting one-dimensional hinge (pointlike corner) modes, see Figs.~\ref{fig:2nd_TI}-~\ref{fig:DSM-spin-1}, pinned at zero quasienergy by an appropriate antiunitary particle-hole symmetry ($P$). In experiments, such dynamic masses can in principle be realized by applying dynamic strain in the system~\cite{strain:exp}, as well as in driven metamaterials, such as acoustic lattices~\cite{bailezhang:acousitcFloquet}. Here we focus on the high-frequency regime, where different Floquet zones remain decoupled~\cite{berg}, and dynamic hinge and corner modes appear at its center. In the future, we will explore the medium- and low-frequency regimes, where these modes can also be found at the Floquet zone boundaries at quasienergies $\pm \omega/2$~\cite{anomalousmode:comment}.

\emph{Acknowledgments}. T.N. thanks MPIPKS, Dresden, for providing the computational facilities. V.J. acknowledges the support of the Swedish Research Council (VR 2019-04735). B.R. was supported by a start-up grant from Lehigh University.


\begin{thebibliography}{}
\bibitem{codimension} The codimension $d_c=d-d_B$, where $d(d_B)$ is the dimensionality of the system (boundary mode). Thus for the corner and hinge modes with $d_B=0$ and $d_B=1$, respectively, $d_c=d$ and $d-1$.
\bibitem{terminology} Often in literature, the unitary (antiunitary) spectral symmetry is referred as chiral (particle-hole) symmetry.
\bibitem{comment1} Other eight $(\pm \sqrt{2}, 0, \pm 1)$ and $(0,\pm \sqrt{2}, \pm 1)$ directions, along which $V_2$ vanishes, are already gapped by $V_1$.
\bibitem{anomalousmode:comment} If one drives a first-order topological phase (insulator or semimetal) with symmetry breaking Wilson-Dirac mass, the system does not support anomalous corner or hinge mode. However, when periodically driven by the first-order mass ($m$), the system tends to accommodate anomalous corner and hinge modes, when accompanied by the symmetry breaking masses ($H^{\rm stat}_{\rm SOTI}$ and $H^{\rm stat}_{\rm TOTI}$, for example), at least in the medium and low frequencies.
\end{thebibliography}
\end{document}